\documentclass{article}
\usepackage{ijcai25}

\usepackage{times}
\usepackage{soul}
\usepackage{url}
\usepackage[hidelinks]{hyperref}
\usepackage[utf8]{inputenc}
\usepackage[small]{caption}
\usepackage{graphicx}
\usepackage{amsmath}
\usepackage{amsthm}
\usepackage{booktabs}
\usepackage{algorithm}
\usepackage{algorithmic}
\usepackage[switch]{lineno}
\usepackage{lipsum}
\usepackage{amssymb}
\usepackage{mathrsfs}
\usepackage{tikz}
\usepackage{multirow}
\usepackage{makecell}
\usepackage{colortbl}
\usepackage{pifont}
\usepackage{subfigure}

\newcommand{\blackcircle}[1]{
    \begin{tikzpicture}[baseline=(X.base)]
        \node[draw, circle, fill=black, inner sep=1pt] (X) {\textbf{\textcolor{white}{#1}}};
    \end{tikzpicture}
}

\title{ECC-SNN: Cost-Effective Edge-Cloud Collaboration for Spiking Neural Networks}

\author{
Di Yu$^1$\and
Changze Lv$^2$\and
Xin Du$^1$\thanks{Corresponding Authors: Xin Du and Shuiguang Deng.}\and
Linshan Jiang$^3$\and
Wentao Tong$^1$\and
Zhenyu Liao$^1$\and
Xiaoqing Zheng$^2$\And
Shuiguang Deng$^1$\\
\affiliations
$^1$Zhejiang University\\
$^2$Fudan University\\
$^3$National University of Singapore\\
\emails
yudi2023@zju.edu.cn,
czlv22@m.fudan.edu.cn,
xindu@zju.edu.cn, linshan@nus.edu.sg \\
\{toldzera, liaozy\}@zju.edu.cn
zhengxq@fudan.edu.cn,
dengsg@zju.edu.cn
}

\begin{document}

\maketitle

\begin{abstract}
    Most edge-cloud collaboration frameworks rely on the substantial computational and storage capabilities of cloud-based artificial neural networks (ANNs). However, this reliance results in significant communication overhead between edge devices and the cloud and high computational energy consumption, especially when applied to resource-constrained edge devices. To address these challenges, we propose ECC-SNN, a novel edge-cloud collaboration framework incorporating energy-efficient spiking neural networks (SNNs) to offload more computational workload from the cloud to the edge, thereby improving cost-effectiveness and reducing reliance on the cloud. ECC-SNN employs a joint training approach that integrates ANN and SNN models, enabling edge devices to leverage knowledge from cloud models for enhanced performance while reducing energy consumption and processing latency. Furthermore, ECC-SNN features an on-device incremental learning algorithm that enables edge models to continuously adapt to dynamic environments, reducing the communication overhead and resource consumption associated with frequent cloud update requests.  Extensive experimental results on four datasets demonstrate that ECC-SNN improves accuracy by 4.15\%, reduces average energy consumption by 79.4\%, and lowers average processing latency by 39.1\%.
    
\end{abstract}

\section{Introduction}
Collaborative edge-cloud computing \cite{wang2024end} leverages the extensive computational power of the cloud alongside the low-latency benefits of edge computing, enhancing data processing efficiency and real-time performance in Internet of Things (IoT) scenarios.
As Spiking Neural Networks (SNNs) have emerged as a promising alternative to Artificial Neural Networks (ANNs), offering greater energy efficiency and responsiveness for on-device intelligence \cite{ijcai2024p596,lvefficient}, there is a growing trend toward replacing edge-side models in conventional edge-cloud frameworks with SNNs.
However, this substitution, while advantageous, introduces several significant challenges.



First, SNNs often underperform compared to ANNs of similar scale in ANN-optimized tasks, such as conventional RGB image classification \cite{deng2020rethinking}. 
This performance gap is primarily attributed to accumulating gradient errors \cite{deng2023surrogate} during the back-propagation through time (BPTT) training process with surrogate functions \cite{wu2019direct}. While SNNs effectively handle most inputs, they struggle with corner cases in the long tail of data distributions—scenarios where fine-tuned ANNs typically excel.

Second, IoT devices continuously collect vast amounts of heterogeneous data from dynamic environments. Uploading all new data to the cloud for updates and retrieving revised models introduces significant communication overhead, increased latency, and excessive energy consumption \cite{long2021complexity}. When numerous devices request updates simultaneously, the cloud server's responsiveness and efficiency are further compromised. These limitations underscore the urgent need for efficient on-device incremental learning algorithms explicitly tailored for SNNs to enable adaptive and scalable edge-cloud collaboration frameworks.

To tackle these challenges, we propose a novel and adaptive edge-cloud collaboration framework, termed \textit{ECC-SNN}\footnote{https://github.com/AmazingDD/ECC-SNN}, which integrates the high inference accuracy of cloud-based ANN models with the responsiveness and energy efficiency of edge-based SNN models.
In ECC-SNN, we first establish a robust edge SNN backbone model, enhanced through a joint ANN-SNN training approach \cite{xu2023constructing}. Unlike previous approaches \cite{jiang2023unified,wang2023new} that improve SNN image classification accuracy by integrating ANN-inspired structures, often requiring similar architectures for both ANN and SNN models, our approach uses knowledge distillation to transfer rich information from pre-trained cloud-based ANNs to edge SNNs with arbitrary architectures, bypassing structural constraints. This method accelerates the training process while minimizing memory consumption during the \textit{Setup} stage of ECC-SNN.

The inference \textit{Execution} process of ECC-SNN also involves both the edge and the cloud with an ambiguity-aware strategy \cite{huang2021integrated}. During inference, inputs with low inference confidence—often representing corner cases or unlearned samples \cite{li2021appealnet}—are regarded as ambiguous or `difficult' cases and are offloaded to the robust cloud-based ANN model for reevaluation. In contrast, those with high confidence, identified as `simple' samples, are directly processed \cite{bolukbasi2017adaptive} by the edge SNN model. Hence, ECC-SNN can achieve a superior cost-accuracy trade-off when performing conventional image classification inference tasks compared to edge- or cloud-only solutions.

Once the edge device completes its allocated tasks, it transitions to the offline \textit{Update} stage (e.g., during charging at the base station), where it performs adaptive on-device incremental learning (IL). During this \textit{Execution} stage, the edge SNN learns from previously ambiguous data and corresponding logits provided by the cloud-based ANN. 
Given the resource constraints of edge devices, implementing complex IL methods \cite{xiao2024hebbian,zhou2022memo} that demand significant memory and computational resources may not be feasible. Therefore, ECC-SNN adopts a simple yet effective self-distillation IL approach \cite{li2017learning}, specifically designed to mitigate \textit{catastrophic forgetting}. Through local updates, the edge SNN improves its performance on difficult inputs, enabling it to continuously adapt to dynamic environmental changes without additional computational overhead from the cloud.

To assess the effectiveness of ECC-SNN, we conduct extensive experiments across four image classification datasets. Results demonstrate that ECC-SNN can adaptively develop a more robust edge-based SNN model in dynamic IoT environments that performs with greater confidence while effectively offloading computational and communication costs from the cloud server. To conclude, our main contributions are summarized as follows:
\begin{itemize}
    \item We propose \textbf{ECC-SNN}, the first edge-cloud collaboration framework to integrate cloud-based ANNs with edge-based SNNs. This novel integration leverages the complementary strengths of ANNs (high inference accuracy) and SNNs (energy efficiency and low latency), marking the first exploration of such a collaboration strategy.  
    \item To address the dynamic IoT environmental changes, we propose an on-device incremental learning method that enhances the adaptability of edge SNN models in ECC-SNN. Furthermore, we incorporate an ambiguity filtering strategy in edge-cloud co-inference, ensuring stable accuracy performance while significantly reducing energy consumption. 
    \item Extensive experiments on \textbf{four} diverse datasets demonstrate that ECC-SNN significantly outperforms standalone edge-based SNNs and cloud-based ANNs, achieving an average accuracy improvement of 4.15\%, a 79.4\% reduction in energy consumption, and a 39.1\% decrease in processing latency.
\end{itemize}

\section{Related Work}
\paragraph{Collaborative Edge-Cloud Computing.}
Most edge-cloud collaboration frameworks heavily rely on computationally powerful servers for execution, with edge devices deploying only small models to handle basic tasks \cite{zhang2024dvfo}. 
For example, ECSeg \cite{zhouecseg} conducts an edge-cloud switched image segmentation system in autonomous vehicles with different delay tolerances.
\cite{li2024distributed,long2021complexity} investigate distributed inference through fine-grained model partitioning, enabling collaborative computation between servers and IoT devices.
In GKT \cite{yao2024gkt}, edge models generate final responses with guidance prompts from larger language models in the cloud. 
In these works, the functionality of edge models remains highly limited due to resource constraints, motivating the exploration of more efficient alternatives to enhance edge capabilities.

\paragraph{Spiking Neural Networks on the Edge.} 
The low power cost and rapid response capabilities of SNNs \cite{maass1997networks} align with resource-constrained edge scenarios, enabling the implementation of more extensive functionalities at the edge. EC-SNN \cite{ijcai2024p596} employs a device collaboration method to distribute deep SNNs at the edge. 
Tr-Spiking-YOLO \cite{yuan2024trainable} is conducted on Jetson devices to implement low-latency objective detection tasks. 
\cite{zhu2024autonomous} proposes to apply an end-to-end training strategy for SNNs on autonomous driving scenarios. 
However, existing SNN applications on edge devices primarily focus on efficient deployment. As devices operate and continuously collect new data, shifts in data distribution can lead to performance degradation, driving the introduction of on-device incremental learning methods tailored for SNNs.

\paragraph{On-Device Incremental Learning.} 
Several studies have been conducted to enhance IL on edge devices, alleviating the catastrophic forgetting problem caused by dynamic environmental changes. Rehearsal-based IL methods \cite{zhou2022memo} achieve favorable performance-cost trade-offs for on-device scenarios \cite{kwon2021exploring}. 
\cite{paissan2024structured} introduces latent replay with sparse weight updates to reduce the learning cost.
\cite{ma2023cost,lee2022carm} employ systematic optimization methods to enhance the efficiency of limited exemplar buffers. However, these studies are primarily ANN-oriented. For on-device incremental training with SNNs, the training cost of BPTT scales with the number of time steps, making allocating additional memory for exemplars impractical. Hence, we focus on devising exemplar-free incremental learning methods tailored for SNNs.


\section{Preliminary}

\subsection{Spiking Neural Networks}

Due to the energy efficiency, SNNs are suitable for deployment on resource-constrained edge devices to implement inference tasks \cite{ijcai2024p596}. In this study, we build the edge SNN architecture with a promising neuron model termed \textit{Leaky Integrate-and-Fire} (LIF) \cite{maass1997networks}, described by a series of discrete-time expressions:
\begin{align} \label{eq:update-mem}
     U(t) & = (1 - \tau)\cdot H(t-1) + \tau \cdot I(t) \\ 
     \label{eq:heaviside}
     O(t) &= \Theta(U(t) - \overline{V}) \\ 
     H(t) &= U(t)\cdot (1-O(t)) + V_r\cdot O(t)
\end{align}
where $\tau, \overline{V}$, and $V_r$ denote the decay factor, threshold, and reset potential. $I(t)$ is the spatial input at time step $t$, $U(t)$ and $H(t)$ are the corresponding membrane potential before and after firing. Additionally, $\Theta(\cdot)$ is the Heaviside step function determining whether a spike is generated. In this study, we adopt BPTT \cite{dampfhoffer2023backpropagation} to train SNNs with surrogate gradients \cite{wu2019direct}. 

\subsection{Prior Probability Distribution Drift}

The rapid influx of data (e.g., a sequence of data stream $\mathcal{D}=\{\mathcal{D}_1,...,\mathcal{D}_N \}$) in real-world applications frequently leads to distribution drifts driven by evolving environments \cite{souza2020challenges}. Let the features and labels of $\mathcal{D}_n$ be denoted as $\mathbf{X}_n$ and $Y_n$, respectively, where $n\in \{1, ..., N\}$. 
\textit{Prior probability drift} arises specifically in problems where the features depend on the labels  (i.e., $\mathcal{Y} \rightarrow \mathcal{X}$), characterized by $P(Y_i|\mathbf{X}_i) = P(Y_j|\mathbf{X}_j)$ while $P(Y_i) \neq P(Y_j)$. 
This drift encapsulates critical challenges in adapting models to dynamic data distributions and is prevalent across various domains \cite{zhou2024class,su2023towards,wang2022transtab}, among which \textit{class-incremental learning} represents a typical scenario for this shift \cite{masana2022class}.
In this study, we primarily investigate the foundational capability of ECC-SNN to tackle prior probability drifts, laying the groundwork for extending its application to more complex drift scenarios. 

\section{Methodology}

\subsection{Problem Statement}
\label{sec:problem}

Image classification is one of the conventional cognitive services provided by SNNs \cite{yao2024spike,shi2024spikingresformer,qiu2024gated}.
Although the performance of SNNs on RGB-based image classification tasks still lags behind that of ANNs \cite{xu2023constructing}, their unique energy efficiency makes them more suitable for deployment on resource-constrained edge devices in practical applications tackling these tasks.

Considering a robot vacuum cleaner equipped with an SNN-based image classifier as the edge device for obstacle avoidance, the input image $\mathbf{x}\in \mathcal{X}$ is the data collected from the cleaner's camera, and its corresponding label $y\in \mathcal{Y}$ (e.g., table, sofa, or pets) might be a random variable drawn with a joint data distribution $P(\mathbf{x}, y)$, in which $\mathcal{Y}=\{1, ..., K \}$ and $K$ is the number of labels.
Due to the dynamic and non-ideal conditions in which inputs are collected, objects in the images may not always behave predictably (e.g., \textit{corner cases} like a black cat might hide in a dark area, and new incoming images with \textit{distribution drifts}). 
Such scenarios could lead to incorrect predictions and decisions by the locally equipped SNN model in the cleaner. In this context, relying solely on edge devices to perform all inferences locally may lead to unsatisfactory accuracy and potentially adverse outcomes. 

\subsection{Edge-Cloud Collaboration Framework}
An edge-cloud collaborative framework \cite{li2021appealnet} can tackle the problem mentioned in section~\ref{sec:problem}. 
We first assume two discriminative models, i.e., one is a computationally intensive, high-performance large ANN-based model $f_0$ operating in the cloud with considerable resource utilization, and the other is a smaller, energy-efficient SNN-based model $f_1$ with limited capacity deployed at the edge. Both are trained on a dataset drawn \textit{i.i.d.} from $P(\mathbf{x},y)$, denoted by $f_0: \mathcal{X} \rightarrow \mathcal{Y}$ and $f_1: \mathcal{X} \rightarrow \mathcal{Y}$, respectively. 

During the inference stage, a soft scoring function (a.k.a. \textit{Filter}) $s(z|\mathbf{x}) \in [0, 1]$ with $z\in \{0, 1\}$ is applied to determine whether the small SNN $f_1$ is qualified to classify current input $\mathbf{x}$ or the input $\mathbf{x}$ should be off-loaded to the robust cloud-based ANN for processing $f_0$. Specifically, we assign label $1$ to $z$ if $s(z|\mathbf{x})$ is below some threshold $\delta$ and label it $0$ otherwise, with the expectation that the filter can effectively upload \textit{hard} inputs with low-confidence inferences from edge SNN to the cloud. 
Hence, the final output of an edge-cloud collaboration framework $(f_0, f_1, s)$ w.r.t a specific input $\mathbf{x}$ is:
\begin{align}
    (f_0,f_1,s)(\mathbf{x})=
    \begin{cases}
        f_1(\mathbf{x}), & \text{if } s(1|\mathbf{x}) \leq \delta \\
        f_0(\mathbf{x}), & \text{o.w.}
    \end{cases}
\end{align}
and the performance of each model $f_z$ can be evaluated by $\mathcal{L}(f_z(\mathbf{x}), y)$, where $\mathcal{L}$ could be a \textit{cross-entropy} loss for image classification tasks. Hence, the overall expected loss of the edge-cloud collaboration framework can be calculated :
\begin{align} \label{eq:obj-ecc}
\begin{aligned}
    & \underset{f_0,f_1,s}{\mathrm{min}} \mathbb{E}_{P(\mathbf{x},y)}\mathbb{E}_{s(z|\mathbf{x})}[\mathcal{L}(f_z(\mathbf{x}), y)] \\
    & \text{s.t. }  \mathbb{E}_{P(\mathbf{x},y)}\mathbb{E}_{s(z|\mathbf{x})}[\mathcal{C}(f_z, s, \mathbf{x})] \leq b  
\end{aligned}
\end{align}
$\mathcal{C}(f_z,s,\mathbf{x})$ refers to the total cost of using a specific model for inference, e.g., energy consumption, inference latency, etc.
Equation~(\ref{eq:obj-ecc}) seeks to minimize the expected loss of models in the edge-cloud collaborative system within a specified cost budget, and its objective can be further extended as follows:
\begin{align} \label{eq:obj-ex}
    \mathbb{E}_{P(\mathbf{x},y)}[s(1|\mathbf{x})\mathcal{L}(f_1(\mathbf{x}), y) + (1-s(1|\mathbf{x}))\mathcal{L}(f_0(\mathbf{x}), y) ] 
\end{align}
 In most resource-constrained edge applications, a particular budget constraint $b$ should be established within the framework to ensure the feasibility of system operation.

However, the cloud-based ANN $f_0$ might be highly complex in achieving state-of-the-art (SOTA) performance, and collaboratively optimizing $f_0$ and $f_1$ could lead to slow convergence. To reduce complexity, we assume a machine learning service vendor at a remote data center might provide a fixed, pre-trained complex ANN $f_0$ with high accuracy performance and $f_0$ serves as an oracle function $\mathcal{O}(\cdot)$ to the small SNN $f_1$ during the optimization phase, i.e., it can correctly classify each input from $P(\mathbf{x}, y)$. Therefore, the loss term $\mathcal{L}(f_0(\mathbf{x}), y)=\mathcal{L}(\mathcal{O}(\mathbf{x}), y)$ in Equation~(\ref{eq:obj-ex}) turns zero, as the outputs of the oracle function consistently correspond to the ground-truth values. Equation~(\ref{eq:obj-ecc}) can then be simplified as:
\begin{align} \label{eq:obj-simple}
\begin{aligned}
    & \mathbb{E}_{P(\mathbf{x},y)}[s(1|\mathbf{x})\mathcal{L}(f_1(\mathbf{x}), y)] \\
    & \text{s.t. } \mathbb{E}_{P(\mathbf{x},y)}\mathbb{E}_{s(z|\mathbf{x})}[\mathcal{C}(f_z, s, \mathbf{x})] \leq b
\end{aligned}
\end{align}

\subsection{System Design}

\begin{figure*}[!t]
\centering
\includegraphics[width=\textwidth]{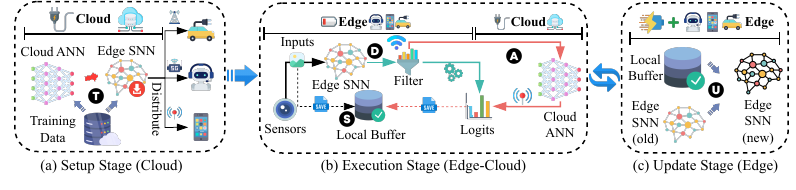}  %
\vspace{-0.2in}
\caption{Overview of the proposed ECC-SNN. In all three stages, the cloud-based ANN model directly or indirectly supports the edge SNN model in preparation, inference, and adaptive updates.}
\vspace{-0.1in}
\label{fig:overview}
\end{figure*}

Figure~\ref{fig:overview} outlines the entire workflow of the proposed \textbf{E}dge-\textbf{C}loud \textbf{C}ollaboration framework for on-device \textbf{S}piking \textbf{N}eural \textbf{N}etwork applications (ECC-SNN for brevity).
The ECC-SNN framework comprises three primary stages. The first \textit{Setup} stage is executed on the cloud, during which a well-trained cloud-based ANN is utilized as a teacher model for jointly optimizing the initial, small-scale SNN, which will be deployed on edge devices to conduct inference tasks. In the \textit{Execution} stage, the edge SNN will request assistance from the cloud-based ANN to classify ambiguous inputs collected from sensor data, which it cannot confidently infer. These inputs, which are hard for the current SNN to classify, will subsequently be labeled by the cloud-based ANN and stored in the local device buffer. These labeled `hard' samples will be utilized for on-device updates in the \textit{Update} stage to improve the SNN's performance in the next \textit{Execution} stage.

\subsubsection{Joint Training Approach}

Although surrogate gradient methods enable directly training non-differentiable SNN, converging the training process is still complicated due to the self-accumulating gradient errors \cite{deng2023surrogate}. To alleviate this problem, we adopt a joint training approach (process \blackcircle{T} in Figure~\ref{fig:overview}), which distills the logit information\footnote{Logit refers to the output of the model’s final classification layer} from the fine-tuned cloud-based ANN to the rough SNN model when prepared at \textit{Setup} stage. Hence, the loss term in Equation~(\ref{eq:obj-simple}) is extended as:
\begin{align}\label{eq:loss1}
    \mathcal{L}(f_1(\mathbf{x}), y) = \mathcal{L}_{ce} + \lambda_1\mathcal{L}_{logit}
\end{align}
where $\mathcal{L}_{ce}$ is the cross-entropy loss to help the SNN learn from the samples in task-specific training datasets, and $\mathcal{L}_{logit}$ refers to the KL-divergence that lets the rough SNN learn the prediction distribution of the cloud-based ANN with $\lambda_1$ controls the corresponding weight.

\begin{figure}[!t]
\centering
\includegraphics[width=\columnwidth]{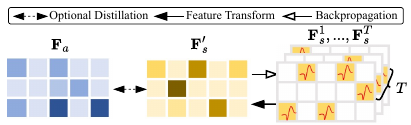} 
\vspace{-0.2in}
\caption{Optional feature distillation in the joint training approach. $T$ is the number of time steps of the features out of the last SNN layer overlapping with the corresponding ANN layer.}
\vspace{-0.1in}
\label{fig:feat-loss}
\end{figure}

This overlap motivates the transfer of feature knowledge between the corresponding parts of the cloud-based ANN and the edge SNN. However, these overlapping intermediate features differ in data format between ANNs and SNNs: in ANNs, features are represented in floating-point format, whereas in SNNs, they are conveyed as time-varying spike trains. To solve the disparity, we introduce external linear modules \cite{qiu2024self} for aligning features, mapping them to the same feature space, as depicted in Figure~\ref{fig:feat-loss}. Hence, for the last overlapping layer $i$ in edge SNN, we define the feature alignment loss as $\mathcal{L}_{align}^i=|| \mathbf{F}_{a, i} - \mathbf{F}_{s,i}' ||_2$ to measure the similarity between feature-pair, in which:
\begin{align}
    \mathbf{F}_{s,i}'=\mathrm{BatchNorm}(\mathrm{Linear}(\sum_t \mathbf{F}_{s,i}^t))
\end{align}
and $\mathbf{F}_{a,i}\in\mathbb{R}^{N\times D}$ and $\mathbf{F}_{s,i}^t\in\mathbb{R}^{N\times D}$ at time step $t$ represent the corresponding decimal and binary feature matrix for the last overlapping ANN and SNN layer $i$, respectively. In summary, the loss term in Equation~(\ref{eq:obj-simple}) for \textit{overlapping} cases turns to:
\begin{align}\label{eq:loss11}
    \mathcal{L}(f_1(\mathbf{x}), y)= \mathcal{L}_{ce} + \lambda_1\mathcal{L}_{logit} + \lambda_2\mathcal{L}_{align}^i
\end{align}
where $\lambda_2$ controls the weight of feature alignment loss. For \textit{non-overlapping} cases, we remain the loss as Equation~(\ref{eq:loss1}).

\subsubsection{Collaborative Inference Strategy}

\begin{figure}[!t]
\centering
\includegraphics[width=\columnwidth]{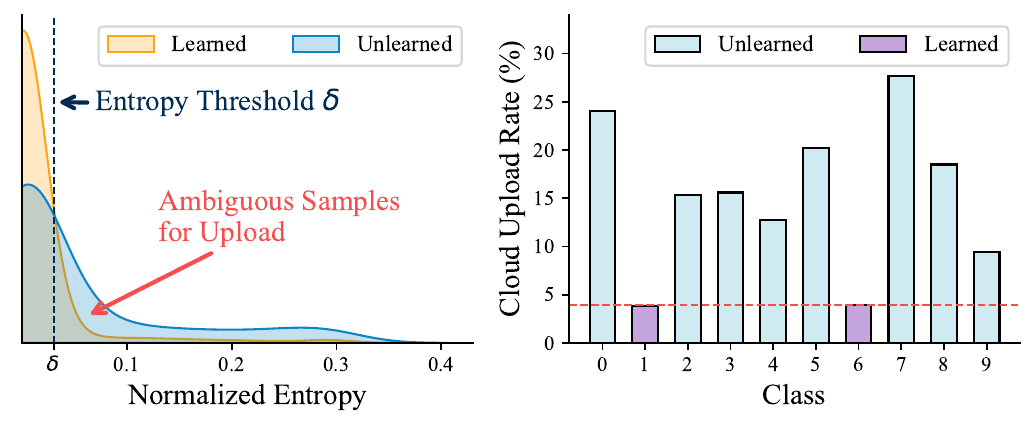} 
\vspace{-0.2in}
\caption{Case Study: a \textit{spiking VGG-9} model learned with CIFAR-10 limited to samples labeled 1 and 6. The entropy distributions and corresponding cloud upload rates for each label are derived from each test sample's model output predictive distribution.}
\vspace{-0.1in}
\label{fig:ed}
\end{figure}

The core of edge-cloud co-inference lies in when to filter inputs that are ambiguous to the current edge model and seek assistance from the cloud. Existing filtering strategies include rule-based \cite{huang2020lightweight} and learning-based \cite{li2021appealnet}. Unlike rule-based approaches, learning-based strategies incur additional computational costs and demand incremental updates as task complexity increases, posing challenges for execution on resource-constrained edge devices. Therefore, \textit{normalized entropy} \cite{huang2021integrated} is employed in ECC-SNN as the filtering criterion $s(1|\textbf{x})$ to determine the cloud upload rate for ambiguous inputs during \textit{Execution} stage. As demonstrated in Figure~\ref{fig:ed}, this metric provides a practical and interpretable measure of inference confidence for the edge SNN model and is defined as follows:
\begin{align}\label{eq:ne}
    s(1|\mathbf{x})=-\sum_{k=1}^K \frac{\sigma(f_{1,k}(\mathbf{x})) \mathrm{log}(\sigma(f_{1,k}(\mathbf{x})))}{\mathrm{log}(K)}
\end{align}
where $\sigma(\cdot)$ denotes the soft-max function converting the output of each edge SNN's classification head $f_{1,k}(\cdot)$ into a probability distribution. To be more specific, the filter in ECC-SNN will compute the corresponding entropy values for each input with the edge SNN (process \blackcircle{D} in Figure~\ref{fig:overview}). Inputs with normalized entropy greater than the pre-set threshold $\delta$ are determined ambiguous samples that require uploading to the cloud for further assistance.

Algorithm~\ref{alg:co-infer} describes how ECC-SNN performs collaborative inference during the \textit{Execution} stage. Each newly collected image from the sensor will first undergo inference using the edge SNN, and the resulting filtering score $s(1|\mathbf{x})$ will be computed to assess the inference confidence. Images with low confidence will be uploaded to the cloud, where the powerful ANN will perform re-inference and return the results. These samples requiring assistance will subsequently use the predicted logits from the cloud-based ANN as their label references, forming local training samples stored in the edge device's local buffer for on-device updates of the SNN model.

\begin{algorithm}[!t]
    \caption{Collaborative Inference in ECC-SNN}
    \label{alg:co-infer}
    \raggedright \textbf{Input}: collected input feature $\mathbf{x}$; filter threshold $\delta$. \\
    \raggedright \textbf{Parameter}: cloud ANN model $f_0$; edge SNN model $f_1$. \\
    \raggedright \textbf{Output}: the inference result $\hat{y}$. \\
    \begin{algorithmic}[1] 
        \STATE $\mathbf{z}\leftarrow f_1(\mathbf{x})$, $\mathbf{p}\leftarrow\sigma(\mathbf{z})$;
        \STATE Compute $s(1|\mathbf{x})$ for $\mathbf{p}$ with Equation~(\ref{eq:ne});
        \IF{$s(1|\mathbf{x}) \leq \delta$}
            \STATE $\hat{y} \leftarrow \mathrm{argmax}_k(\mathbf{p})$, $k\in \{1, ...,K\}$;
        \ELSE
            \STATE \texttt{\# line 7-9: process} \blackcircle{A} \texttt{in Figure~\ref{fig:overview}}
            \STATE Upload $\mathbf{x}$ to the cloud, $\mathbf{z}\leftarrow f_0(\mathbf{x})$;
            \STATE $\hat{y} \leftarrow \mathrm{argmax}_k(\mathbf{z})$, $k\in \{1, ...,K\}$;
            \STATE Broadcast $\hat{y}$ back to the edge device; 
            \STATE \texttt{\# line 11: process} \blackcircle{S} \texttt{in Figure~\ref{fig:overview}}
            \STATE Store sample $(\mathbf{x}, \hat{y})$ in local buffer of the edge device;
        \ENDIF
        \STATE \textbf{return} $\hat{y}$
    \end{algorithmic}
\end{algorithm}

\subsubsection{On-device Incremental Learning Method}

Although the co-inference mechanism in ECC-SNN substantially enhances accuracy by refining the results produced by the edge SNN, it may also introduce additional communication latency and energy overheads. This is particularly true when the edge SNN faces challenges related to environmental mobility, such as data distribution drift \cite{ijcai2024p0909}, inducing numerous ambiguous inferences with low confidence. For instance, a robot may be delivered to a user’s home with default object recognition capabilities. However, it may fail to recognize new, site-specific objects reliably (a.k.a. suffering from \textit{prior probability distribution drift}).

Therefore, IL methods are employed to progressively improve the performance of the edge SNN in handling these ambiguities. (process \blackcircle{U} in Figure~\ref{fig:overview}). Meanwhile, this learning process should be conducted on edge devices rather than on a centralized cloud server during the offline stage (e.g., recharging at the base station). In-situ processing helps avoid transmitting high-volume information over networks, thereby reducing the bandwidth requirements \cite{kukreja2019training}.

Unlike humans who continually learn evolving tasks throughout their lifetime, the edge SNN model suffers from \textit{catastrophic forgetting} problems \cite {li2017learning} when conducting IL.
Although many IL methods can mitigate this issue \cite{xiao2024hebbian,zhou2022memo}, they often require additional memory and computational resources, making them unsuitable for deployment on resource-constrained edge devices.
Therefore, our proposed framework 
will provide a bio-plausible explanation for this method) for conducting on-device IL for the edge SNN, which continually training the edge SNN model with a new loss function during \textit{Update} stage: 
\begin{align}\label{eq:loss3}
    \mathcal{L}(f_1(\mathbf{x}), y) = \mathcal{L}_{new} + \lambda_3\mathcal{L}_{old}
\end{align}
where $\mathcal{L}_{new}$ is analogous to Equation~(\ref{eq:loss1}) but is specifically computed using samples stored in the local buffer, supplemented by logits obtained from the cloud-based ANNs.
$\mathcal{L}_{old}$, evaluated by KL-divergence with the probabilities from the old SNN model $\sigma(f_1^{old}(\mathbf{x}))$ and the new SNN model $\sigma(f_1^{new}(\mathbf{x}))$, prevents forgetting the knowledge of previous tasks by forcing the model to predict similar outputs as the previous SNN model for old task data. The importance of $\mathcal{L}_{old}$ is controlled by a hyper-parameter $\lambda_3$.
The local buffer will be flushed after each incremental update is completed.

\section{Experiments}

\subsection{Experimental Settings}
Following \cite{zhou2024class}, we evaluate the effectiveness of our proposed ECC-SNN using the standard class-incremental learning setting, as it represents a typical form of prior probability distribution drift commonly encountered in real-world applications \cite{diao2023oebench}. We denote the data split setting as 'w/ B-$u$, Inc-$v$,' i.e., the first dataset contains $u$ classes, and each following dataset contains $v$ classes. $u=0$ means the total classes are equally divided into each task. 
By default, we adopt the \textit{spiking VGG-9} model as the SNN deployed on the edge device featuring a neuromorphic chip \cite{ma2024darwin3} while utilizing a widely recognized pre-trained ViT model \textit{vit-base-patch16} as the cloud-based ANN.

\subsection{Performance Evaluation}

\begin{table}[!t]
\centering
\resizebox{\columnwidth}{!}{%
\begin{tabular}{llrrr}
\toprule
Dataset & ANN Arch. & $\bar{\mathcal{A}}_{1,\mathrm{E}}$ (\%) & $\bar{\mathcal{A}}_{1,\mathrm{C}}$ (\%) & $\bar{\mathcal{A}}_{1,\mathrm{EC}}$ (\%) \\ 
\midrule
  \multirow{3}{*}{CIFAR-10} & VGG-16$^\dagger$ & 86.75 & 93.41 & \cellcolor{gray!20}(\textcolor{red}{$\uparrow$ 3.82}) 90.57 \\
 & ResNet-34 & 86.75 & 95.23 & \cellcolor{gray!20}(\textcolor{red}{$\uparrow$ 3.79}) 90.54 \\
 & ViT-12 & 86.75 & 98.05 & \cellcolor{gray!20}(\textcolor{red}{$\uparrow$ 3.88 }) 90.63 \\ 
 \cmidrule(l){1-5}
\multirow{3}{*}{CIFAR-100} & VGG-16$^\dagger$ & 83.20 & 89.10 & \cellcolor{gray!20}(\textcolor{red}{$\uparrow$ 2.25}) 85.45 \\
 & ResNet-34 & 83.20 & 91.50 & \cellcolor{gray!20} (\textcolor{red}{$\uparrow$ 2.15}) 85.35 \\
 & ViT-12 & 83.20 & 94.75 & \cellcolor{gray!20}(\textcolor{red}{$\uparrow$ 2.30}) 85.50 \\ 
 \cmidrule(l){1-5}
 \multirow{3}{*}{Caltech} & VGG-16$^\dagger$ & 94.23 & 98.04 & \cellcolor{gray!20}(\textcolor{red}{$\uparrow$ 0.76}) 95.04 \\
 & ResNet-34 & 94.23 & 98.39 & \cellcolor{gray!20}(\textcolor{red}{$\uparrow$ 0.61}) 94.84 \\
 & ViT-12 & 94.23 & 99.19 & \cellcolor{gray!20}(\textcolor{red}{$\uparrow$ 0.76}) 95.04 \\ 
 \cmidrule(l){1-5}
\multirow{3}{*}{Tiny-ImageNet} & VGG-16$^\dagger$ & 62.05 & 69.50 & \cellcolor{gray!20}(\textcolor{red}{$\uparrow$ 2.75}) 64.80 \\
 & ResNet-34 & 62.05 & 75.50 & \cellcolor{gray!20}(\textcolor{red}{$\uparrow$ 4.85}) 66.90 \\
 & ViT-12 & 62.05 & 89.50 & \cellcolor{gray!20}(\textcolor{red}{$\uparrow$ 6.55}) 68.60 \\ 
 \bottomrule
\end{tabular}%
}
\caption{Average accuracy performance at Task 1 ($\bar{\mathcal{A}}_1$) of edge-side \textit{spiking VGG-9} w.r.t. various pre-trained cloud-based ANN architecture (Arch.) pairs across different RGB-based image datasets. Footnotes `$\mathrm{E}$'/`$\mathrm{C}$'/`$\mathrm{EC}$' represent the edge SNN, cloud-based ANN, and ECC-SNN. ANN Arch. with `$\dagger$' indicates the overlapping case in which the dimensions of their early layers are identical to those of the paired SNN. All results are averaged across three random seeds. 
}
\label{tab:overall-acc-base-vgg}
\end{table}

\paragraph{Test Accuracy.} 
Denote the average Top-1 accuracy ($\bar{\mathcal{A}}_n$) after the $n$-th task as $\bar{\mathcal{A}}_n = \frac{1}{n} \sum_{m=1}^n a_{n, m}$, where $a_{n,m}\in [0, 1]$ is the accuracy of task $m$ after learning task $n$ ($m \leq n$).
We first evaluate the effectiveness of the proposed joint training approach for accuracy improvement at \textit{Setup} stage. As listed in Table~\ref{tab:overall-acc-base-vgg}, we compare the performance of this approach with that of standalone training approaches for edge SNN and cloud-based ANN models, which demonstrate that the joint training design in ECC-SNN can achieve an average accuracy improvement of 2.87\% for edge SNNs compared to the standalone direct training approach. 

\begin{table}[!t]
\centering
\resizebox{0.75\columnwidth}{!}{%
\begin{tabular}{llrr}
\toprule
$\mathcal{L}_{align}^i$ & $\mathcal{L}_{logit}$ & CIFAR-100 & Tiny-ImageNet \\ \midrule
\ding{55} & \ding{55} & 83.20 & 62.05 \\
\ding{51} & \ding{55}  & (\textcolor{red}{$\uparrow$ 0.75}) 83.95 & (\textcolor{red}{$\uparrow$ 1.35}) 63.40 \\
\ding{55} & \ding{51} & (\textcolor{red}{$\uparrow$ 0.95}) 84.15 & (\textcolor{red}{$\uparrow$ 2.10}) 64.15 \\ 
\midrule
\rowcolor{gray!20}
\ding{51} & \ding{51} & (\textcolor{red}{$\uparrow$ 2.25}) 85.45 & (\textcolor{red}{$\uparrow$ 2.75}) 64.80 \\ 
\bottomrule
\end{tabular}%
}
\caption{Ablation results of the proposed regularization terms for the overlapping VGG cases. 
}
\label{tab:abl-acc}
\end{table}

The enhancement of SNN performance varies depending on the choice of the pre-trained ANN model. 
Although VGG16 is less complex than other pre-trained ANN models like ResNet and ViT, it can still serve as an effective teacher model to assist SNN convergence on simpler tasks such as CIFAR and Caltech, and the final performance gap is even less than 0.1\%.
However, we must acknowledge that in complex tasks like ImageNet, the performance limitations of VGG-16 itself constrain its ability to provide substantial guidance to Spiking VGG-9. As a result, the final accuracy improvement falls short of the improvement achieved with those complex ANNs.
Observations in Table~\ref{tab:abl-acc} elaborate on the contribution of each proposed regularization term in those VGG-structured overlapping cases, from which we conclude that both terms demonstrate effectiveness for the final accuracy, with
$\mathcal{L}_{logit}$ being the dominant factor. In addition, their combination can further enhance the learning performance across different datasets.

\begin{figure}[!t]
\centering
\includegraphics[width=\columnwidth]{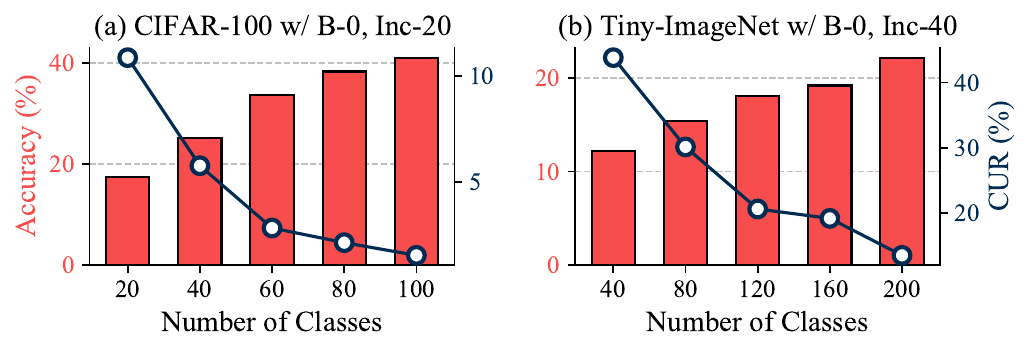} 
\vspace{-0.2in}
\caption{Changing patterns of accuracy performance and CUR as
more classes learned, with a fixed filtering threshold $\delta$=0.3.}
\vspace{-0.1in}
\label{fig:acc-cur}
\end{figure}

The adaptive update strategy in ECC-SNN is designed to continuously enhance the predictive capabilities of the edge SNN model, thereby offloading as much computational burden and communication cost from the cloud-based ANN model as possible.
We define the cloud upload rate (CUR) \cite{li2021appealnet} to represent the proportion of ambiguous inputs uploaded to the cloud for assistance, representing the communication overhead of co-inference and denoted as:
\begin{align}
    \text{CUR}= \mathbb{E}_{P(\mathbf{x}, y)} \left[ \mathbb{I}(s(1|\mathbf{x})>\delta) \right]
\end{align}
where $\mathbb{I}(\cdot)$ is the indicator function. When using the full test dataset as a simulation of a real-world environment to evaluate the proposed system, tendencies in Figure ~\ref{fig:acc-cur} depict that the model's accuracy consistently improves with incremental updates. As the edge SNN model gains higher inference confidence in classifying, it will reduce the reliance on the complicated cloud-based ANN model to recognize ambiguous `difficult' inputs, thereby diminishing CUR.

\begin{figure}[!t]
\centering
\includegraphics[width=\columnwidth]{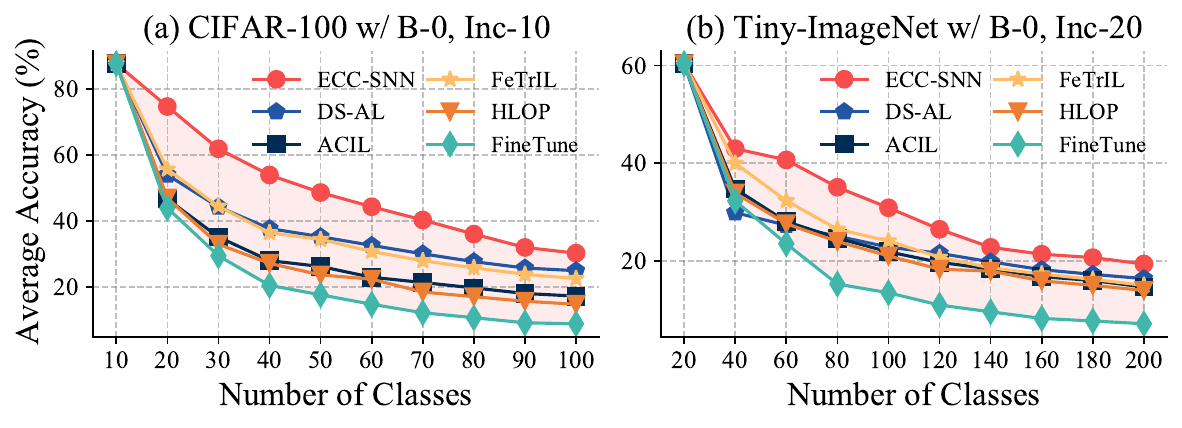} 
\vspace{-0.2in}
\caption{Average accuracy of edge SNN w.r.t. different IL methods.}
\vspace{-0.1in}
\label{fig:il-acc-cmp}
\end{figure}

Figure~\ref{fig:il-acc-cmp} compares various SOTA IL methods proposed recently with the adaptive update approach in our ECC-SNN. 
Note that all methods experience a degradation in $\bar{\mathcal{A}}$ due to catastrophic forgetting as the number of classes increases. 
In ANN-oriented IL tasks, freezing the backbone weights effectively preserves the existing knowledge. However, in SNNs, the weights represent synaptic connections, and freezing them completely restricts the SNN's ability to continue learning.
Therefore, ECC-SNN is more effective at mitigating this impact than other methods across various scenarios, achieving an improvement of 5.32\% over the second-best method on CIFAR-100 and 2.98\% on ImageNet, respectively.

\paragraph{Energy Cost.}
\begin{figure}[!t]
  \centering
  \includegraphics[width=\columnwidth]{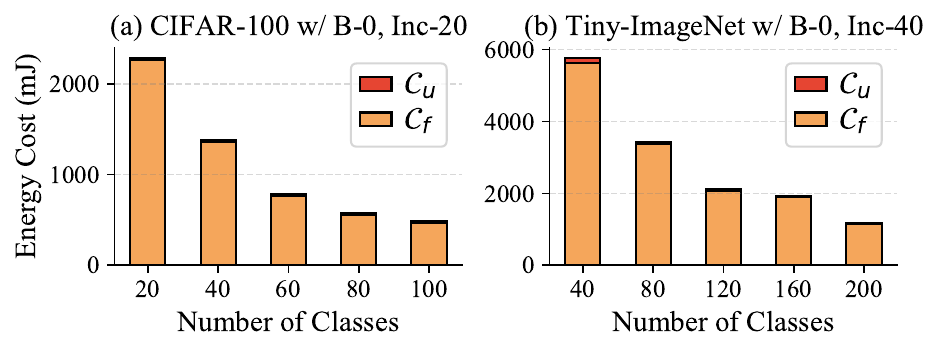} 
  \vspace{-0.2in}
  \caption{Average theoretical energy cost (mJ) per input for inference as more classes learned in ECC-SNN, with a fixed filtering threshold $\delta$=0.3. The communication cost $\mathcal{C}_u$ is negligible compared to the computation cost $\mathcal{C}_f$.}
  \vspace{-0.1in}
  \label{fig:energy}
\end{figure}

The energy cost of ECC-SNN during the \textit{Execution} stage contains two main components: communication overheads $\mathcal{C}_u$ and selected computational cost $\mathcal{C}_f$
. 
Figure~\ref{fig:energy} introduces changing cost patterns as tasks evolve, indicating that (1) The computational cost of the cloud-based ANN model constitutes the majority of the total energy consumption. (2) As the edge model is continuously updated, its inference capability gradually improves, reducing the reliance on the cloud-based model. This enhancement leverages the energy efficiency of SNNs, reducing the average energy consumption for inference. (3) The communication overhead becomes non-negligible as the input size increases. For example, it contributes 2.2\% of the total cost at the first task in the ImageNet scenario.

\begin{figure}[!t]
  \centering
\includegraphics[width=\columnwidth]{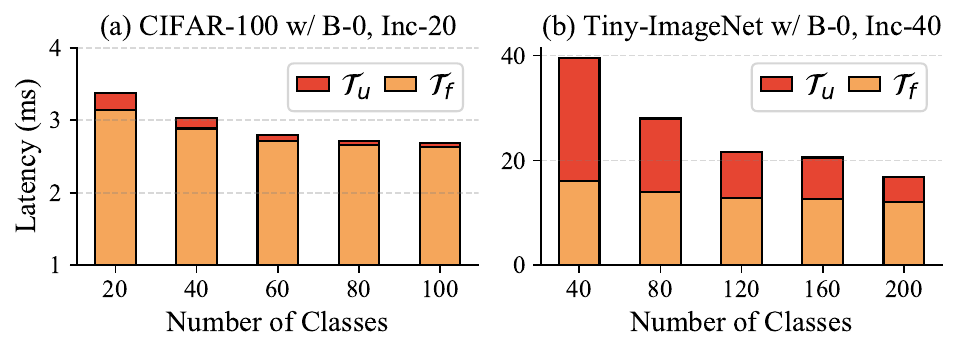} 
  \vspace{-0.2in}
  \caption{Average inference latency (ms) per input in ECC-SNN as more classes learned, with a fixed filtering threshold $\delta$=0.3.}
  \vspace{-0.1in}
  \label{fig:latency}
\end{figure}

\paragraph{Latency.} The inference latency of ECC-SNN also consists of communication latency $\mathcal{T}_u$ and model computation latency $\mathcal{T}_f$.
Figure~\ref{fig:latency} introduces the latency patterns of ECC-SNN when inferring one sample for different datasets as tasks evolve. It can be intuitively concluded that (1) As the capability of the edge SNN improves, reliance on the powerful ANN model on the cloud diminishes, reducing the frequency of request transmissions and the latency associated with waiting for inference results. 
(2) Simple CIFAR-100 inputs are more likely to be processed directly on the edge rather than uploaded to the cloud. Hence, the average computational latency acceleration per input is 9.07\% lower than that in ImageNet.
(3) When the input size is large, running spiking VGG-9 on a neuromorphic chip is only 21.9\% faster than running ViT-12 on a server GPU. Therefore, reducing the frequency of requesting assistance from the cloud cannot diminish the total computation latency but can accelerate the communication latency by around 79.7\%. Therefore, when evaluating the effectiveness of the ECC-SNN system, communication latency and model inference latency should be considered equally important metrics.

\subsection{Sensitive Analysis}

\begin{figure}[!t]
\centering
\includegraphics[width=\columnwidth]{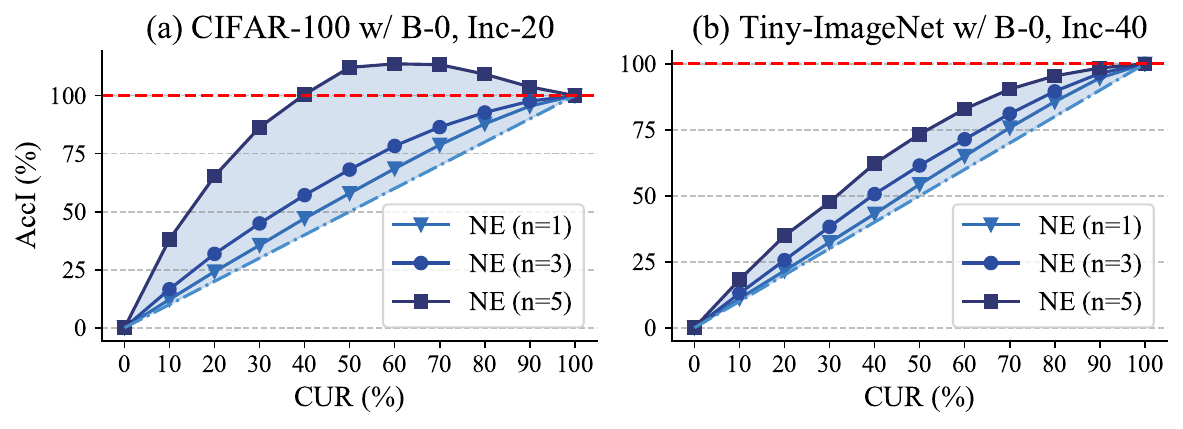} 
\vspace{-0.2in}
\caption{Accuracy improvement ($\mathrm{AccI}$) w.r.t CUR (\%) with different learned classes at current task.}
\vspace{-0.1in}
\label{fig:filters}
\end{figure}

Following \cite{li2021appealnet}, we define the relative accuracy improvement ($\mathrm{AccI}$) to measure the accuracy improvement of ECC-SNN $\mathcal{A}_{(f_0, f_1, s)}$ compared to the standalone SNN $\mathcal{A}_{f_1}$ deployed at the edge, normalized by the accuracy gap between the cloud-based ANN $\mathcal{A}_{f_0}$ and the edge SNN $\mathcal{A}_{f_1}$. 
As shown in Figure~\ref{fig:filters}, the normalized entropy (NE) strategy in ECC-SNN demonstrates its effectiveness by accurately identifying ambiguous inputs and improving the utility of requesting assistance from the cloud model, compared to random uploading. 
Meanwhile, it is evident that as the edge model's predictive capabilities improve, the marginal benefit of requesting assistance from the cloud gradually diminishes. 
A surprising observation is that ECC-SNN can improve accuracy in simple classification tasks. For example, the overall accuracy of ECC-SNN can exceed that of the standalone cloud-based ANN model (red dashed line) by up to approximately 13.3\% in task 5 of CIFAR-100 when the CUR is in the range of [40, 100].
This is because, although the edge SNN initially exhibits lower accuracy across the entire dataset, it is continuously optimized with incoming data from different tasks to minimize overall expected loss. As a result, it is likely to correctly predict a subset of inputs that the cloud-based ANN model fails to classify correctly.
However, for the complex ImageNet dataset, the accuracy gap between the cloud and edge models is too large to enable accuracy boosting. Under such circumstances, the edge model can continuously benefit from the cloud, allowing a better trade-off between accuracy and cost within ECC-SNN.

\section{Conclusion}

In this study, we propose ECC-SNN, a cost-effective and efficient edge-cloud collaborative framework designed for SNN-based classifiers.
By employing the joint training approach and adaptive on-device incremental learning with the assistance of a powerful ANN model on the cloud server, the SNN model in ECC-SNN gains enhanced predictive capability compared to the standalone edge SNN, significantly reducing both energy costs and inference latency as the system operates in different dynamic IoT scenarios.

\section*{Acknowledgments}
The work of this paper is supported by the National Key Research and Development Program of China under Grant 2022YFB4500100, the National Natural Science Foundation of China under Grant 62125206, the
Zhejiang Provincial Natural Science Foundation of China under 
Grant No. LD24F020014, the National Key Research and Development Program of China No. 2024YDLN0005, and the Regional Innovation and Development Joint Fund of the National Natural Science Foundation of China No. U22A6001.

\bibliographystyle{named}
\bibliography{ijcai25}

\end{document}